\documentclass{ctr_summer}

\usepackage{ctrfont}

\usepackage[dvips]{graphicx}
\usepackage{psfrag}
\usepackage{array}
\usepackage{natbib}
\usepackage{amsmath,rotate}
\usepackage{subfigure}

\newcolumntype{M}[1]{>{\centering\arraybackslash}m{#1}}
\title{
Modeling of the subgrid-scale term of the filtered magnetic field transport equation
}
\shorttitle{Modeling of the SGS term of the filtered magnetic field transport equation}

\author{G. Balarac\footnote{LEGI  (Grenoble-INP, UJF, CNRS), France}, A.~G. Kosovichev\footnote{HELP, Stanford University, USA}, O. Brugi\`ere$\dagger$, A.~A. Wray\footnote{NASA Ames Research Center, USA} and N.~N. Mansour$\P$.}

\shortauthor{G. Balarac, A. Kosovichev, O. Brugi\`ere, A. Wray and N. Mansour}

\begin{document}

\maketitle

Accurate subgrid-scale turbulence models are needed to perform realistic numerical magnetohydrodynamic (MHD) simulations of the subsurface flows of the Sun. To perform large-eddy simulations (LES) of turbulent MHD flows, three unknown terms have to be modeled. As a first step, this work proposes to use {\it a priori} tests to measure the accuracy of various models proposed to predict the SGS term appearing in the transport equation of the filtered magnetic field. It is proposed to evaluate the SGS model accuracy in term of ``structural'' and ``functional'' performance, i.e. the model capacity to
locally approximate the unknown term and to reproduce its energetic action, respectively. From our tests, it appears that a mixed model based on the scale-similarity model has better performance.

\section{Introduction}



Significant progress towards quantitative understanding of the Sun and predictive capabilities for solar activity
and space weather requires large-scale, integrated modeling of the physical conditions in subsurface layers
of the Sun. Realistic numerical magnetohydrodynamic (MHD) simulations of the subsurface flows and magnetic structures have
become achievable because of the development of fast supercomputer systems and efficient parallel computer
codes. These simulations are extremely important for understanding the complicated physics of the upper
turbulent convective boundary layer of the Sun. The dynamics of this layer is critical for understanding of the
formation of magnetic regions on the Sun and their activity. This layer is also a source of solar oscillations.
The wave excitation and propagation properties change dramatically in strong field regions. Their modeling
and investigation are very important for local helioseismology and helioseismic data analysis.
Accurate Large Eddy Simulations (LES) of the subsurface dynamics depend on the development of specific and accurate subgrid-scale turbulence models.
These models have to provide a realistic description of effects of small-scale unresolved turbulence, which are particularly important for studying wave excitation.

In LES of MHD flows, to know the filtered velocity, $\bar{u}_i$, and magnetic, $\bar{b}_i$, fields, the filtered MHD equations, expressed in Alfven-speed units, have to be solved,
\begin{equation} \label{ns1}
\dfrac{\partial \bar{u}_i}{\partial t} + \dfrac{\partial \bar{u}_i \bar{u}_j}{\partial x_j} = - \dfrac{1}{\rho} \dfrac{\partial \bar{p}}{\partial x_i} + \dfrac{\partial }{\partial x_j} \left( \nu \dfrac{\partial u_i}{\partial x_j}- \tau^{u}_{ij}+\tau^{b}_{ij}\right) + \dfrac{\partial \bar{b}_i \bar{b}_j}{\partial x_j},
\end{equation}
\begin{equation} \label{ns2}
\dfrac{\partial \bar{b}_i}{\partial t} + \dfrac{\partial  }{\partial x_j} (\bar{b}_i \bar{u}_j - \bar{b}_j \bar{u}_i) =  \dfrac{\partial }{\partial x_j} \left( D \dfrac{\partial \bar{b}_i}{\partial x_j} - \tau^{ub}_{ij}\right) \mbox{ and }
\end{equation}
\begin{equation} \label{ns3}
\dfrac{\partial \bar{b}_i}{\partial x_i} = \dfrac{\partial \bar{u}_i}{\partial x_i} = 0,
\end{equation}
where $D$ is magnetic diffusivity, $\nu$ is kinematic viscosity, $\bar{p}$ is filtered total pressure. In these equations,   $\tau^{u}_{ij} = \overline{u_i u_j} - \bar{u}_i \bar{u}_j$, $\tau^b_{ij}=\overline{b_i b_j} - \bar{b}_j \bar{b}_i$, and  $\tau^{ub}_{ij} = \overline{b_i u_j} - \bar{b}_i\bar{u}_j - \left( \overline{u_i b_j} - \bar{u}_i \bar{b}_j \right)$ are subgrid-scale (SGS) tensors. In a LES, these SGS tensors cannot explicitly be determined but are estimated via SGS models assuming relationships with resolved quantities.
Various works have addressed the modeling of the three SGS tensors of LES of MHD flows \cite[]{Yoshizawa:1990,Theobald:1994,Muller:2002,Miki:2008}. In this work, we propose to investigate the performance of the models proposed for $\tau^{ub}_{ij}$ from {\it a priori} tests. The goal is to distinguish the modeling error due to the models of $\tau^{ub}_{ij}$ from the other modeling errors. The modeling performances are evaluated as ``structural'' and ``functional'' performances.

\section{Modeling of $\tau^{ub}_{ij}$ and performance measurement}

\subsection{Available SGS models}

The modeling of the unknown SGS term, $\tau^{ub}_{ij}$, appearing in the filtered transport equation of the magnetic field has been addressed in various ways in the past. First, an approach based on the definition of an eddy magnetic diffusivity, $D_t$, has been proposed, leading to the general gradient-diffusion model expression
\begin{equation} \label{eddy_dif}
\tau^{ub}_{ij} = - 2 D_t \bar{J}_{ij},
\end{equation}
with $\bar{J}_{ij} = \frac{1}{2}\left( \frac{\partial \bar{b}_j}{\partial x_i} - \frac{\partial \bar{b}_i}{\partial x_i}\right)$, the filtered magnetic rotation tensor. Various definitions of $D_t$ are available in the literature. Various extensions of the Smagorinsky model \cite[]{Smag1963} have been proposed. First, \cite{Yoshizawa:1987} defined the eddy magnetic diffusivity as
\begin{equation} \label{Yoshi}
D_t = C_{\lambda} \Delta^2 \left( \frac{1}{2}C_{\nu}\bar{S}^2_{ij}+C_{\lambda}\bar{j}^2_i\right)^{1/2},
\end{equation}
with $\bar{S}_{ij} = \frac{1}{2}\left( \frac{\partial \bar{u}_j}{\partial x_i} + \frac{\partial \bar{u}_i}{\partial x_i}\right)$, the resolved velocity rate-of-strain tensor and $\vec{j}=\vec{\nabla} \times \vec{b}$, the current density. Assuming a local equilibrium between production and dissipation, \cite{Hamba:2010} determine the constant values as $C_{\lambda}=\frac{5}{7}C_{\nu}$ and $C_{\nu}=0.046$. \cite{Theobald:1994} propose to define $D_t$ only with the current density norm,
\begin{equation} \label{theobald}
D_t = C_1 \Delta^2 |\bar{\vec{j}}|.
\end{equation}
\cite{Muller:2002} use this model by computing dynamically the model coefficient as usually done for LES of hydrodynamic turbulence\cite[]{Lilly_PFA_1992,Germano_POF}.
In the same paper, \cite{Muller:2002} define a new eddy magnetic diffusivity based on the cross helicity dissipation,
\begin{equation} \label{crosshelicity}
D_t = C_2 \Delta^2 |\bar{\vec{j}} . \bar{\vec{\omega}}|^{1/2},
\end{equation}
with $\vec{\omega}$ the vorticity vector. The model coefficient is also computed dynamically.

Another approach to model $\tau^{ub}_{ij}$ is based on the filtering operation itself. For example, a Taylor series expansion of a filtered product, $\overline{fg}$ (where $f$ and $g$ are quantities describing magnetic or flow fields), can be given for a Gaussian filter as (see \cite{Balarac_POF_2008} for details)
\begin{equation} \label{bye}
\overline{fg} =  \bar{f}\bar{g}
              + \frac{\Delta^2}{12} \frac{\partial \bar{f}}{\partial x_i} \frac{\partial \bar{g}}{\partial x_i}
              + \frac{\Delta^4}{288} \frac{\partial^2 \bar{f}}{\partial x_i \partial x_j}
                                     \frac{\partial^2 \bar{g}}{\partial x_i \partial x_j} \\
              + \frac{\Delta^6}{10368} \frac{\partial^3 \bar{f}}{\partial x_i \partial x_j \partial x_k}
                                       \frac{\partial^3 \bar{g}}{\partial x_i \partial x_j \partial x_k}
              +... \mbox{  .}
\end{equation}
Keeping only the first term, a gradient model for $\tau^{ub}_{ij}$ can be written as
\begin{equation} \label{gm}
\tau^{ub}_{ij} = \frac{\Delta^2}{12} \left(\frac{\partial \bar{b}_i}{\partial x_k} \frac{\partial \bar{u}_j}{\partial x_k} - \frac{\partial \bar{b}_j}{\partial x_k} \frac{\partial \bar{u}_i}{\partial x_k}\right).
\end{equation}
Another model can be constructed from the scale-similarity hypothesis proposed by \cite{Bardina}. The main idea is to assume that the statistical structure of the tensors constructed on the basis of the subgrid scales is similar to that of
their equivalents evaluated on the basis of the smallest resolved scales. From this assumption, the scale-similarity model for $\tau^{ub}_{ij}$ can be written as
\begin{equation} \label{ssm}
\tau^{ub}_{ij} = \overline{\bar{b}_i\bar{u}_j} -  \overline{\bar{b}_j\bar{u}_i} - \left( \bar{\bar{b}}_i \bar{\bar{u}}_j - \bar{\bar{b}}_j \bar{\bar{u}}_i \right).
\end{equation}
However, these types of models are known to not provide enough energy transfer
between grid scale (GS) and subgrid-scale (SGS), leading to unstable simulations when they
is used to close the filtered Navier-Stokes equations \cite[]{Vreman:1997}. To avoid
this unstable behavior, a mixed model can be proposed. As proposed by \cite{Clark_JFM_1979} for the Navier-Stokes equations, the mixed model consists to add an eddy magnetic diffusivity approach to the gradient or the scale-similarity model. Using the model (\ref{theobald}), the mixed-gradient model is thus written as
\begin{equation}  \label{mgm}
\tau^{ub}_{ij} =  C_3 \Delta^2 |\bar{\vec{j}}|\bar{J}_{ij}+\frac{\Delta^2}{12} \left(\frac{\partial \bar{b}_i}{\partial x_k} \frac{\partial \bar{u}_j}{\partial x_k} - \frac{\partial \bar{b}_j}{\partial x_k} \frac{\partial \bar{u}_i}{\partial x_k}\right),
\end{equation}
whereas the mixed-scale-similarity model is written as
\begin{equation}  \label{mssm}
\tau^{ub}_{ij} =  C_4 \Delta^2 |\bar{\vec{j}}|\bar{J}_{ij}+ \overline{\bar{b}_i\bar{u}_j} -  \overline{\bar{b}_j\bar{u}_i} - \left( \bar{\bar{b}}_i \bar{\bar{u}}_j - \bar{\bar{b}}_j \bar{\bar{u}}_i \right),
\end{equation}
with $C_3$ and $C_4$ dynamically computed.

Finally, a last approach is inspired from works in RANS context \cite[]{Yoshizawa:1990}. Thus, \cite{Miki:2008} proposed to model $\tau^{ub}_{ij}$ as
\begin{equation}  \label{ef}
\tau^{ub}_{ij} = -2\epsilon_{kij}E_k,
\end{equation}
where $\epsilon_{kij}$ is the Levi-Civita symbol and $E_k$ is the turbulence electromotive force defined as
\begin{equation} \label{mm}
E_k = \alpha \bar{b}_k - \beta \bar{j}_k + \gamma \bar{\omega}_k.
\end{equation}
In this equation, $\alpha$, $\beta$ and $\gamma$ are spatially dependent quantities defined with the kinetic and magnetic subgrid-scale energy, $k^{sgs}=\frac{1}{2}\left(\overline{u_i u_i} -\bar{u}_i\bar{u}_i\right)$ and $k^{sgs,b}=\frac{1}{2}\left(\overline{b_i b_i} -\bar{b}_i\bar{b}_i\right)$, and computing model coefficient locally and dynamically (see \cite{Miki:2008}  for details). Note that this model needs to solve two additional transport equation for $k^{sgs}$ and $k^{sgs,b}$ with additional unknown terms to model.
\\

\begin{table}
\begin{center}
\begin{tabular}{|c|c|c|c|} 
Name  & Equations & Symbol & Remark \\
 & & & \\
Yoshizawa & (\ref{eddy_dif}) and (\ref{Yoshi}) & \circle & 2 constant coefficients \\
Theolbald & (\ref{eddy_dif}) and (\ref{theobald}) & \squarop & 1 dynamic coefficient \\
Cross-helicity & (\ref{eddy_dif}) and (\ref{crosshelicity}) & \diamop & 1 dynamic coefficient \\
Gradient & (\ref{gm}) & \trianopu & 1 constant coefficient \\
Scale-similarity & (\ref{ssm}) &  \trianopd & Explicit filtering \\
Mixed-Gradient & (\ref{mgm}) & \trianopu (dotted line) & 1 dynamic coefficient \\
Mixed-Scale-similarity & (\ref{mssm}) &  \trianopd (dotted line) & 1 dynamic coefficient \\
Miki &  (\ref{ef}) and (\ref{mm}) & \trianopr & 2 additional transport equations \\
\end{tabular}
\end{center}
\label{tablemodel}
\caption{Models for the unknown SGS term $\tau^{ub}_{ij}$}%
\end{table}
Eight SGS models for $\tau^{ub}_{ij}$ presented above have been summarized in table \ref{tablemodel}. Some of these models are very easy to apply (especially without dynamic coefficient computation) but some others need a
non-negligible implement effort and have a non-negligible computational cost. The goal of this work is to be able to determine which model appears as the best way to model $\tau^{ub}_{ij}$. Our analysis is mainly based on {\it a priori} tests using direct numerical simulation (DNS) data. Our {\it a priori} tests are performed to measure two types of performance. Using the distinction between structural and functional models proposed by \cite{sagaut_book}, we propose to measure the SGS model performance in terms of structural and functional performance.

\subsection{Structural performance and optimal estimator}

The structural modeling strategy consists in making the best approximation of the unknown SGS term by constructing it from the knowledge of the structure of small-scales. For example, the gradient model, Eq.(\ref{gm}), is a structural model based on a Taylor series expansion of the filtering operation. This type of model is known to give a good approximation of the unknown term with a high correlation between the unknown term and the model in {\it a priori} test. From this definition, we define the structural performance of a model by its capacity to locally approximate the SGS term to be modeled.

To evaluate the structural performance, the optimal estimation theory is used.  In the framework of optimal estimation theory \cite[]{Deutsch_book}, the models are compared using the notion of an optimal estimator \cite[]{Moreau_POF_2006}. Based on this idea, if a quantity $\tau$ is modeled with a finite set of variables $\phi$, an exact model cannot be guaranteed.
If the exact solution $\tau$ is known, for example from DNS,
the optimal estimator of $\tau$ in terms of the set of variables $\phi$ is given by the expectation of the quantity $\tau$ conditioned on the variables in the set, i.e. $\langle \tau | \phi \rangle$, where the angular brackets indicate statistical averaging over a suitable ensemble. A quadratic error can consequently be defined as the average of the square of the difference at each point between the conditional mean value given by the value of $\phi$ at this point and the exact value of the quantity,
\begin{equation} \label{IE}
\epsilon_{\rm min} = \langle \left(\tau - \langle \tau | \phi \rangle \right)^2 \rangle,
\end{equation}
where $\epsilon_{\rm min}$ is the error. It should be noted that any model formulated using the variable set $\phi$ will introduce an error that is larger than or equal to this minimum error, with the best model formulation producing this minimum error. Consequently, this quadratic error $\epsilon_{\rm min}$ is referred to as the irreducible error. Only a change in the variable set may reduce the magnitude of this error.  In contrast, the total quadratic error is given as
\begin{equation} \label{TE}
\epsilon_{\rm tot} = \langle (\tau - f(\phi))^2 \rangle,
\end{equation}
with $f(\phi)$ the proposed model for $\tau$. The method allows to compare different LES
models by comparing their total errors. The method allows also to evaluate the improvement possibility of a given model without changing of the set of parameters. Indeed,  if the total error  of a given model is much higher than its
irreducible part, improvement can be expected (by modification of the coefficient computation, for example). This method finally allows to find the best set of quantities to model a sub-filter term by comparing their irreducible error.

\subsection{Functional performance}

The functional modeling strategy consists to consider  the action of the subgrid terms on the transported quantity (here, the magnetic field) and not the unknown term itself. It can consist in introducing a dissipative term, for example, that as a similar effect but not necessarily the same structure. Since an adequate mechanism to dissipate magnetic energy from resolved to subgrid scales is essential,  we define the functional performance as the model capacity to lead to the good energy dissipation. The transport equation of the GS magnetic energy, $\bar{k}^{b} = \frac{1}{2}\bar{b}_i \bar{b}_i$, is
$$
\frac{\partial \bar{k}^{b}}{\partial t} + \bar{u}_i\frac{\partial \bar{k}^{b}}{\partial x_i} = \frac{1}{2} D \frac{\partial^2 \bar{k}^{b}}{\partial x_i \partial x_i} - D  \frac{\partial \bar{b}_i}{\partial x_j}\frac{\partial \bar{b}_i}{\partial x_j}
+ \frac{\partial}{\partial x_j} \left( \bar{b}_j \bar{b}_i \bar{u}_i\right) - \bar{u}_i \frac{\partial}{\partial x_j} \left( \bar{b}_j \bar{b}_i \right) - \frac{\partial}{\partial x_j} \left( \bar{b}_i \tau^{ub}_{ij} \right) + \tau^{ub}_{ij} \bar{J}_{ij}.
$$
In this equation, the GS/SGS energy exchanges is due to the SGS dissipation, $\tau^{ub}_{ij} \bar{J}_{ij}$. The functional performance will be then evaluated as the model capacity to well predict this quantity in a statistical sense.

\section{Results}

\subsection{Numerical method}

\begin{figure}
\begin{center}
    \psfrag{x}[cc]{$k$}
    \psfrag{y}[cc]{$E(k)$, $E_B(k)$}
    \includegraphics[width = 0.4\textwidth]{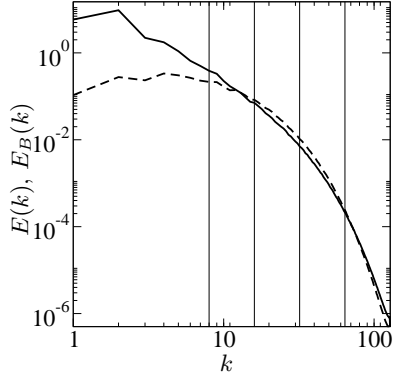}
    \caption{
    Kinetic, $E(k)$, and magnetic, $E_B(k)$, energy spectra:  $E(k)$ \solid; $E_B(k)$ \dashed. The thin vertical lines show the location of some filters used in this work: $\Delta/\Delta x =$2, 4, 8 and 16.
}
    \label{fig1}
\end{center}
\end{figure}
As already explained, to clearly identify the performances of the various modeling strategy of $\tau^{ub}_{ij} $ without taking into account the modeling error of the other SGS unknown terms appearing in the filtered MHD equations (\ref{ns1})-(\ref{ns3}), we performed {\it a priori} tests. {\it A priori} tests are conducted using direct numerical simulation (DNS) data from a forced homogeneous isotropic turbulence computation. A pseudo-spectral code with second-order explicit Runge-Kutta time-advancement is used. The viscous terms are treated exactly. The simulation domain is discretized using $256^3$ grid points on a domain of length $2\pi$.  A classic 3/2 rule is used for dealiasing the non-linear convection term, and statistical stationarity is achieved using a forcing term \cite[]{Alvelius_POF_1999}.  The transport equation of the magnetic field is advanced simultaneously using an identical numerical scheme. First, a hydrodynamic (no magnetic field) case is performed; when the statistically stationary state is obtained, the Reynolds number based on the Taylor microscale is around $100$. The magnetic field is then initialized at small scales with a small amplitude. The magnetic Prandtl number is set to $0.5$. Without external forcing, the magnetic energy grows leading quickly to a new statistical state. The {\it a priori tests} are performed when the flow is statistically stationary.

In the {\it a priori} tests, explicit filter is used to replicate the behavior of the filter
implicitly associated with the discretization in real LES. Two kinds of filter are used, a spectral cutoff
filter to mimic spectral LES and a box filter to mimic LES using centered finite differences \cite[]{Rogallo_ARFM_1984}.
Several different filter sizes have been used such as $ 2 \leq \Delta / \Delta x \leq 16$, where $\Delta$ is the filter width and $\Delta x$ is the grid spacing used in the DNS. The location in wavenumber space of the filters used are displayed in Fig. \ref{fig1}, which shows the kinetic and magnetic  energy spectra.

\subsection{Models performances}

\begin{figure}
\begin{center}
    \psfrag{x}[cc]{$\Delta / \Delta x$}
    \psfrag{y}[cb]{Error}
    \includegraphics[width = 0.4\textwidth]{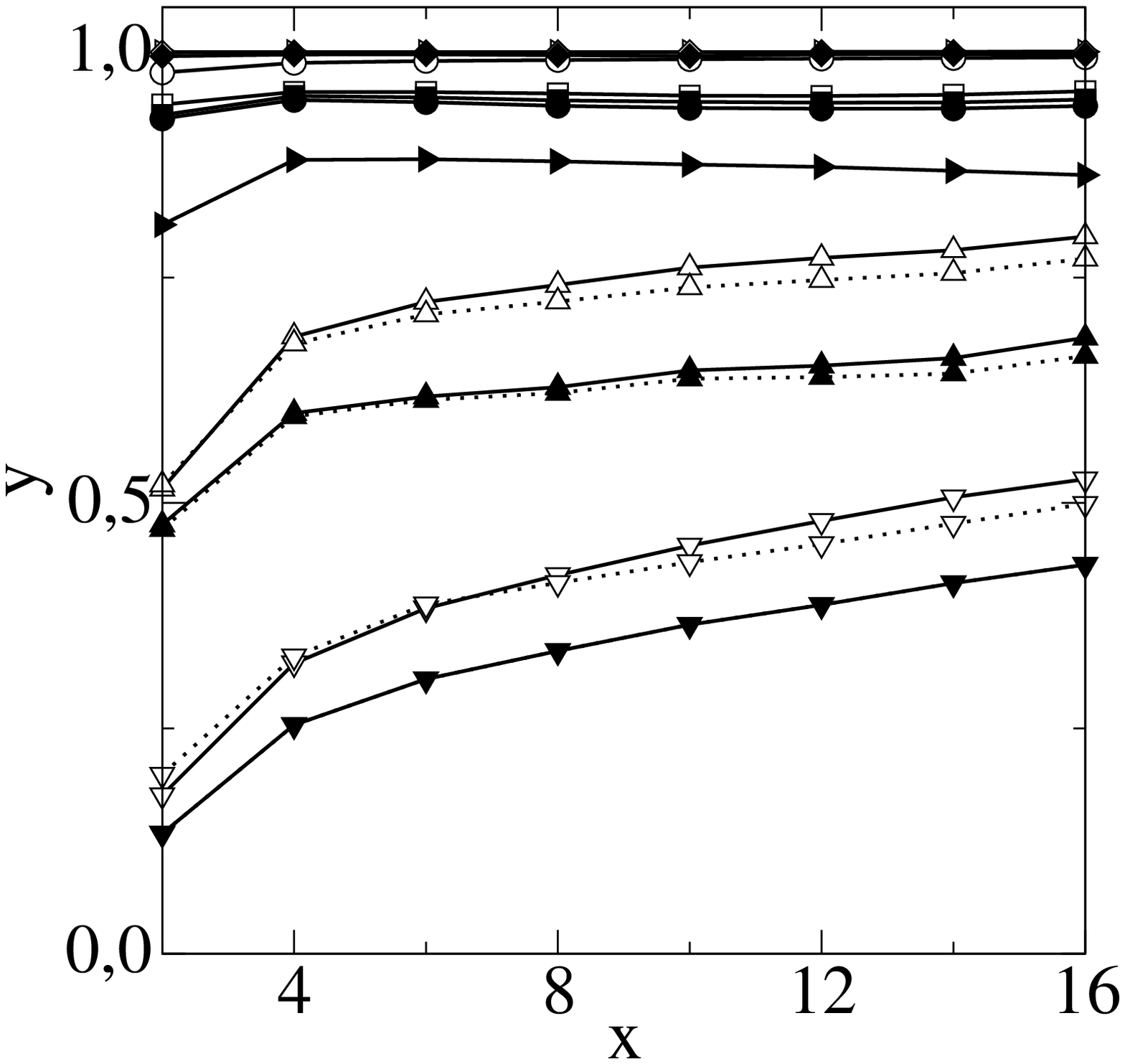}
    \hspace{1cm}
     \includegraphics[width = 0.4\textwidth]{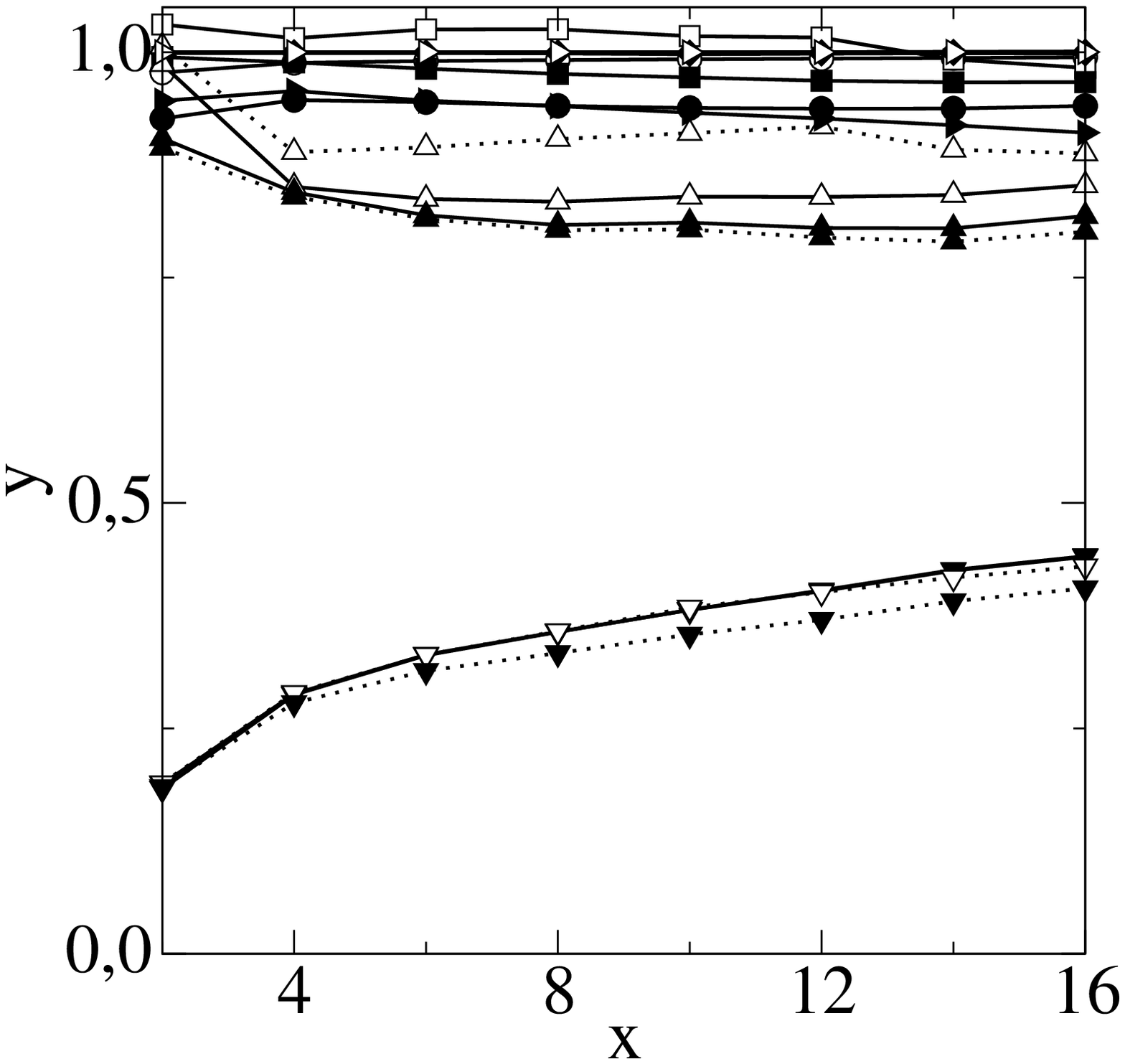}
    \caption{
    Structural performance of models. Evolution of the irreducible and total errors, Eq. (\ref{IE}) and (\ref{TE}), with the filter size for box (left) and spectral cutoff (right) filter. Open and solid symbols are for total and irreducible errors, respectively (see Table \ref{tablemodel} for symbols correspondence).
}
    \label{fig2}
\end{center}
\end{figure}
As already explained, to first evaluate the structural performance of the SGS models, the total error, Eq. (\ref{TE}) for each model is considered and compared with its irreducible error, Eq. (\ref{IE}). Figure \ref{fig2} shows the evolution of the total and irreducible error with the filter size, for the different SGS models and for both box and spectral cutoff filter. On this figure, the error is normalized by the statistical variance of the SGS term.
First conclusions can be addressed. As expected, SGS models based on structural approach (models based on scale-similarity assumption or on Taylor series expansion) lead to the smallest errors. Note that for both filters the models based on the scale-similarity assumption have the smallest errors. In particular for the spectral cutoff filter the errors of the models based on a Taylor series expansion stay high. This is because the spectral cutoff filter leads to a divergent Taylor series, because of its non-localness \cite[]{sagaut_book}. All the other models have irreducible errors higher than the total error model of the models based on structural approach. This shows that a structural improvement of these models can not be expected without adding new quantities in its set of parameters. This allows to illustrate the improvement due to the mixed approach in comparison with the models based only on an eddy diffusivity assumption.

\begin{figure}
\begin{center}
    \psfrag{x}[cc]{$\Delta / \Delta x$}
    \psfrag{y}[cb]{$\langle \tau_{ij}^{ub} \bar{J}_{ij} \rangle$}
    \includegraphics[width = 0.4\textwidth]{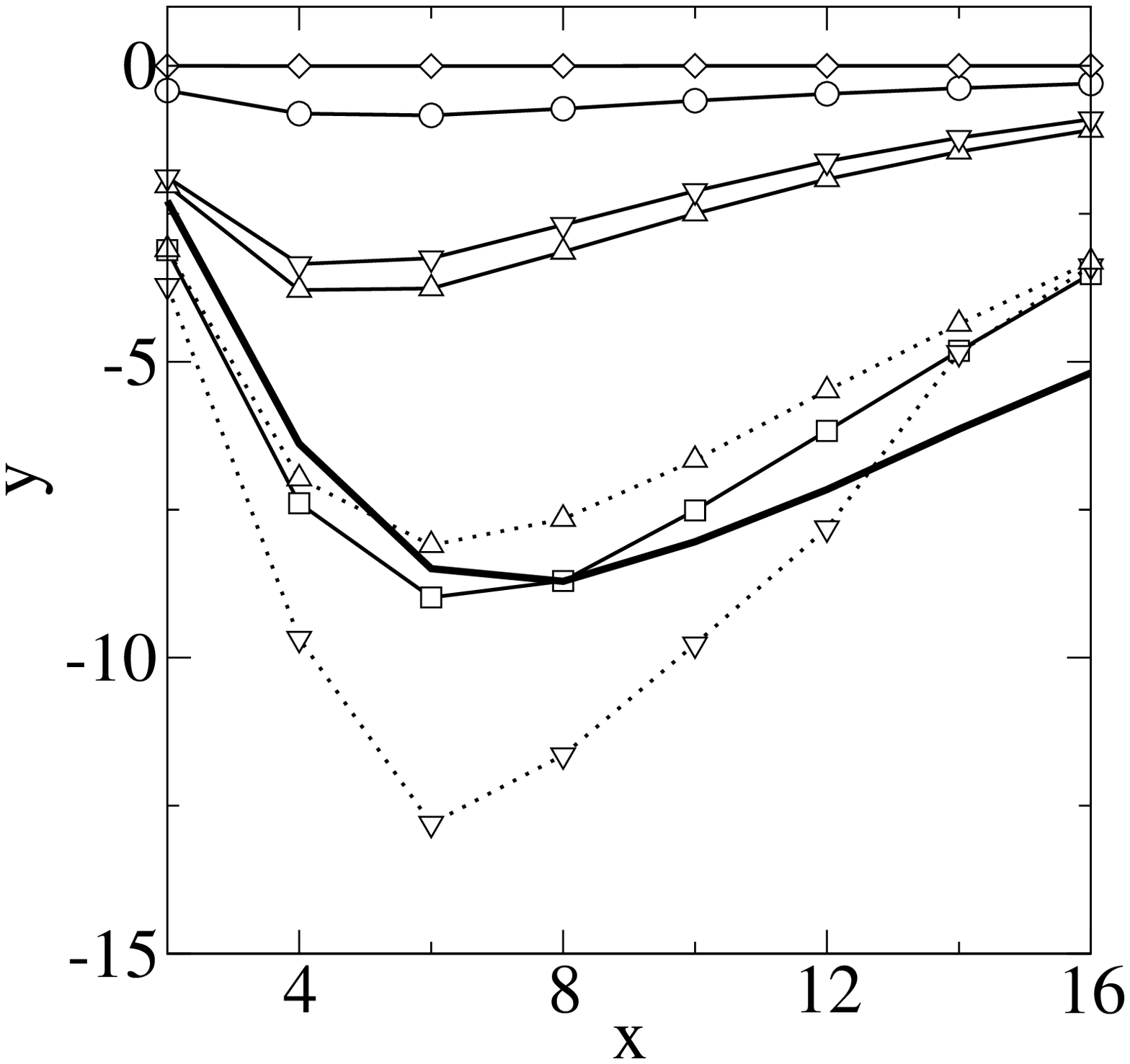}
    \hspace{1cm}
     \includegraphics[width = 0.4\textwidth]{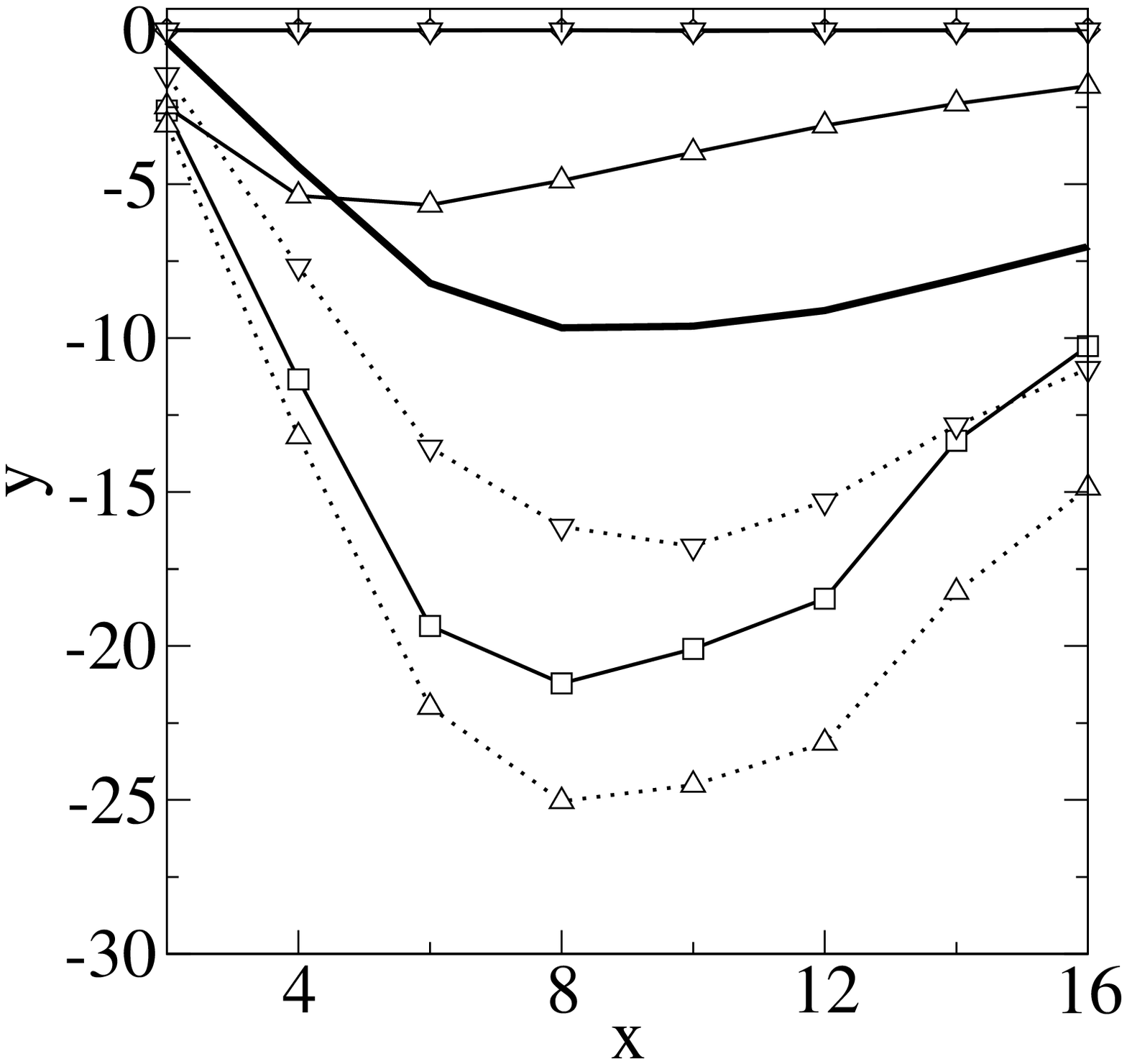}
    \caption{
    Functional performance of models. Evolution of the mean GS/SGS magnetic energy transfer with the filter size for box (left) and spectral cutoff (right) filter (see Table \ref{tablemodel} for symbols correspondence). The thick line is the mean GS/SGS magnetic energy transfer from the filtered DNS.
}
    \label{fig3}
\end{center}
\end{figure}
As second step, the functional performance is now studied from the evolution of the mean GS/SGS magnetic energy transfer, $\langle \tau_{ij}^{ub} \bar{J}_{ij} \rangle$, with the filter size. Figure \ref{fig3} shows the results for box and spectral cutoff filters. First, it is shown that the ``pure'' structural models, i.e. the gradient and scale-similarity models, give not enough transfer in comparison with the DNS results. This is a well-known problem for models based only on structural approach for hydrodynamic LES. Indeed this model as known to not give enough dissipation, leading to unstable simulations. Conversely, Theobald's eddy-diffusivity based model predicts enough GS/SGS transfer and even an over-estimation for the spectral cutoff filter in comparison with the DNS results. Thus, this allows the mixed models to predict enough dissipation. Note that the other eddy-diffusivity based models do not lead to enough GS/SGS transfer to be a good candidate to build a mixed model. In particular, the cross-helicity model leads to no GS/SGS transfer. \cite{Muller:2002} had already shown this property explaining that in LES performed with this SGS model, the lack of GS/SGS magnetic energy transfer is compensated by the transfer between kinetic and magnetic energy caused by the Lorentz force.

From this analysis, the best performing model appears to be the mixed-scale-similarity model. First, for the structural performance, the analysis based on the total and irreducible errors shows that we can not expect an improvement of the other models to have the same performance of this model. Indeed the total error of this model is always smaller than the irreducible error, i.e. the smallest possible error, of the other models. For the functional performance, the eddy-diffusivity part of this model allows to predict enough dissipation to allow the simulation stability. It can be however noted that a possible improvement of this model would be to correct the over-estimation of the GS/SGS transfer.

\subsection{Dynamic procedure at the tensor divergence level}

\begin{figure}
\begin{center}
    \psfrag{x}[cc]{$\Delta / \Delta x$}
   \psfrag{y}[cb]{Error}
    \psfrag{z}[cb]{$\langle \tau_{ij}^{ub} \bar{J}_{ij} \rangle$}
    \includegraphics[width = 0.4\textwidth]{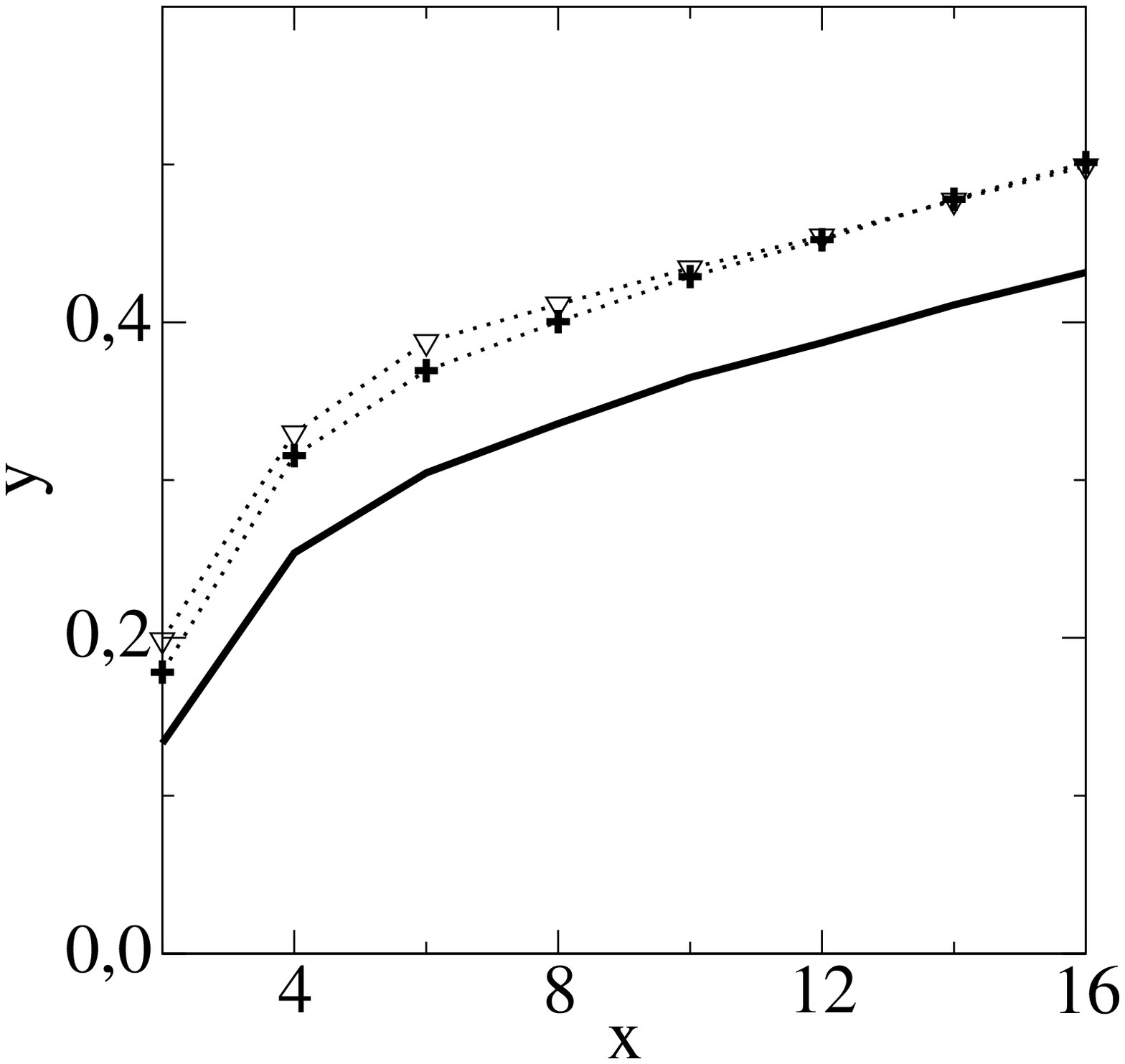}
    \hspace{1cm}
     \includegraphics[width = 0.4\textwidth]{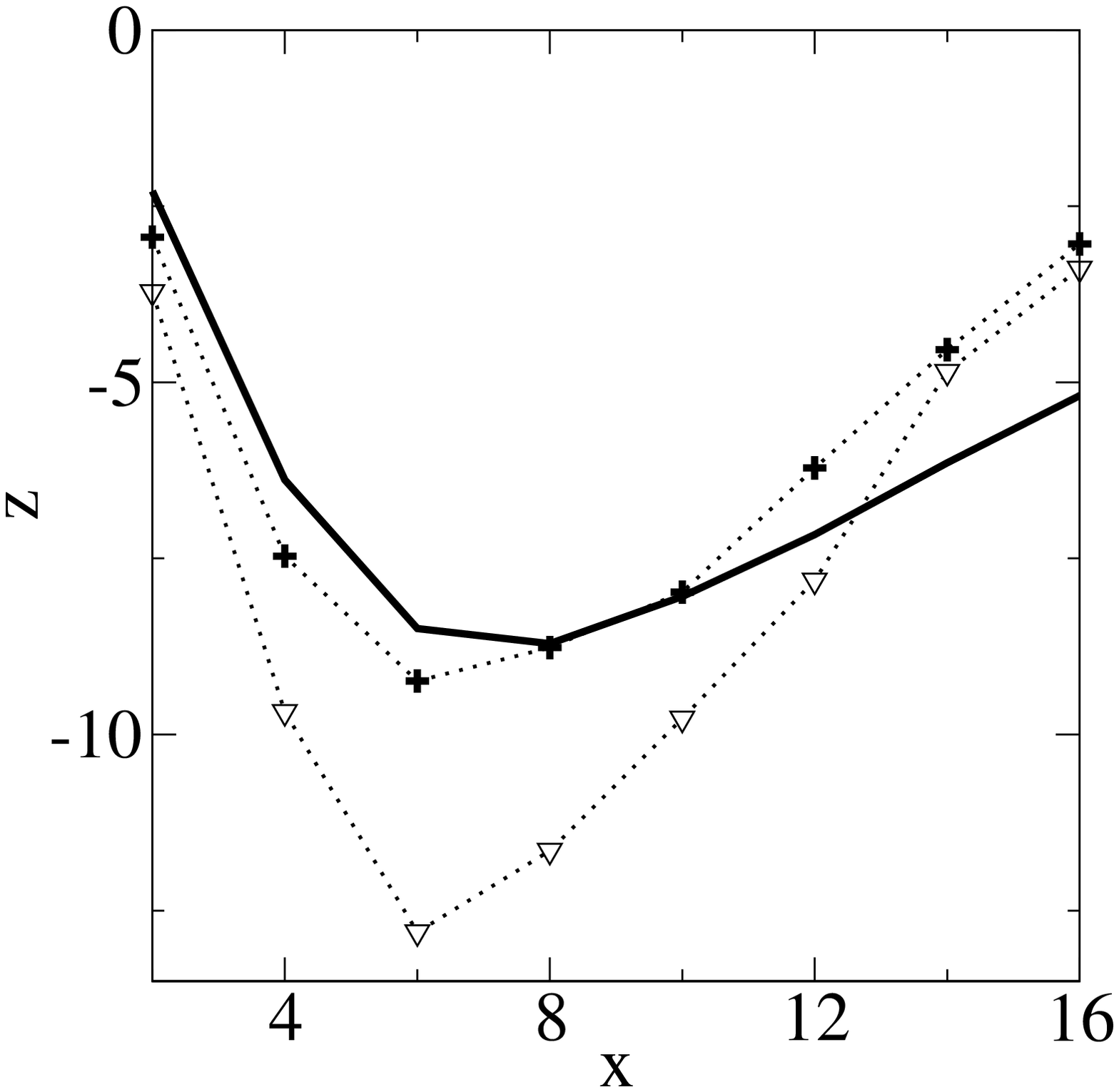}
    \caption{
Comparison of model performance for the mixed-scale-similarity model with the classical (triangle) and with the divergence based (cross) dynamic procedure. Left: evolution of the irreducible (solid line) and total errors with the filter size.
Right: Evolution of the mean GS/SGS magnetic energy transfer with the filter size and comparison with the mean GS/SGS magnetic energy transfer from the filtered DNS (solid line).
}
    \label{fig4}
\end{center}
\end{figure}
In the results above, the mixed-scale-similarity model, Eq. (\ref{mssm}), used a classic dynamic procedure to compute $C_4$. The classic dynamic procedure uses a second (test) filter, denoted $\hat{.}$, of size $\hat{\Delta} = 2 \Delta$. This procedure is  based on the Germano identity \cite[]{Germano_POF},
$$
L^{ub}_{ij} = T^{ub}_{ij} - \hat{\tau}^{ub}_{ij},
$$
where $T^{ub}_{ij} = \widehat{\overline{b_i u_j}}-\hat{\bar{b}}_i \hat{\bar{u}}_j - \left( \widehat{\overline{u_i b_j}}-\hat{\bar{u}}_i \hat{\bar{b}}_j\right)$. Thus,
$L^{ub}_{ij} =  \widehat{\bar{b}_i \bar{u}_j}-\hat{\bar{b}}_i \hat{\bar{u}}_j - \left( \widehat{\bar{u}_i \bar{b}_j}-\hat{\bar{u}}_i \hat{\bar{b}}_j\right)$ can be computed directly from the resolved field. Assuming that, $T^{ub}_{ij}$, which is the subgrid tensor corresponding to the second filtering level, is also modeled with the mixed-scale-similarity model and with the same value of $C_4$. An equation for $C_4$ can then be written from a least squares averaging procedure \cite[]{Lilly_PFA_1992}.
In fact, from the filtered transport equation of the magnetic field, Eq. (\ref{ns2}), it can be noted that only the vector given by the divergence of the tensor, $\partial \tau^{ub}_{ij} / \partial x_j$ has to be known and not  the tensor, $\tau^{ub}_{ij}$, itself. In this sense, \cite{Clark_JFM_1979} explained the efficiency of the Smagorinsky model for the Navier-Stokes equations.  It is shown that the correlation between the Smagorinsky model and the SGS term is weak at the tensor level but higher at the vector level. Thus to improve the prediction of the GS/SGS magnetic energy transfer with the mixed-scale-similarity model, a dynamic computation of the $C_4$ coefficient is tested for $\partial \tau^{ub}_{ij} / \partial x_j$ instead of $ \tau^{ub}_{ij} $. The starting point is thus to use the divergence of the Germano identity
$$
\frac{\partial L^{ub}_{ij}}{\partial x_j} = \frac{\partial T^{ub}_{ij} }{\partial x_j}- \widehat{ \frac{\partial \tau^{ub}_{ij}}{\partial x_j} },
$$
and the same steps of the classical dynamic procedure are then used.
Figure \ref{fig4} shows the structural and functional performance of the mixed-scale-similarity model using a divergence based dynamic procedure. The results are compared with the classic mixed-scale-similarity model. It is first important to note that the structural performance has not deteriorated. The total error of the model stays close to its irreducible error and the error is much smaller than the error of the other models (Fig. \ref{fig2}). Moreover, the functional performance is improved. The new dynamic procedure allows a better prediction of the GS/SGS magnetic energy transfer. The over-prediction observed with the classic dynamic procedure disappears and the transfer predicted by the model is close to the transfer computed from the filtered DNS.

\begin{figure}
\begin{center}
    \psfrag{x}[cc]{time}
   \psfrag{y}[cb]{$\bar{u}_i^2(t)/\bar{u}_j(0)$}
    \psfrag{z}[cb]{$\bar{b}_i^2(t)/\bar{b}_j(0)$}
    \includegraphics[width = 0.4\textwidth]{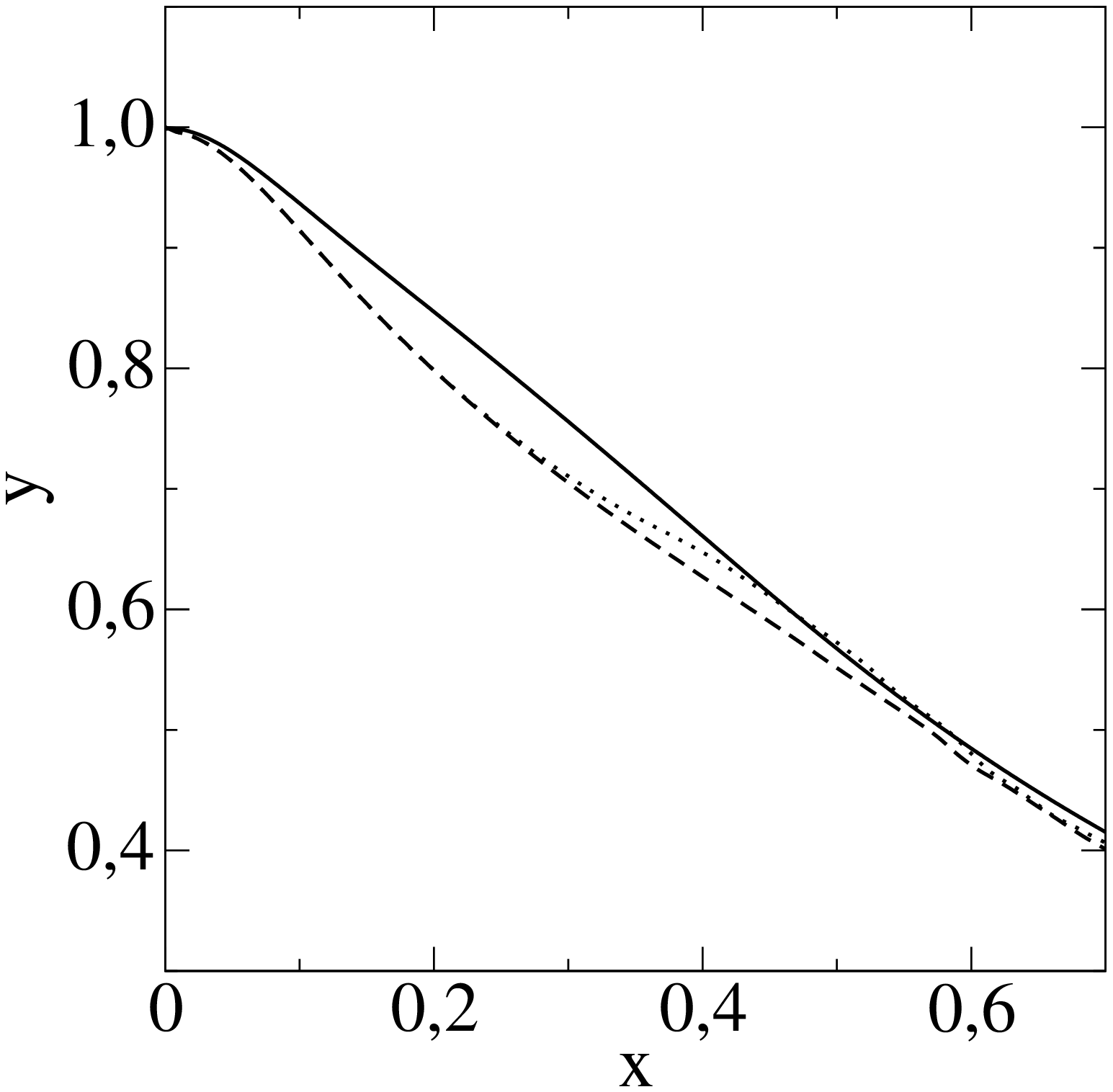}
    \hspace{1cm}
     \includegraphics[width = 0.4\textwidth]{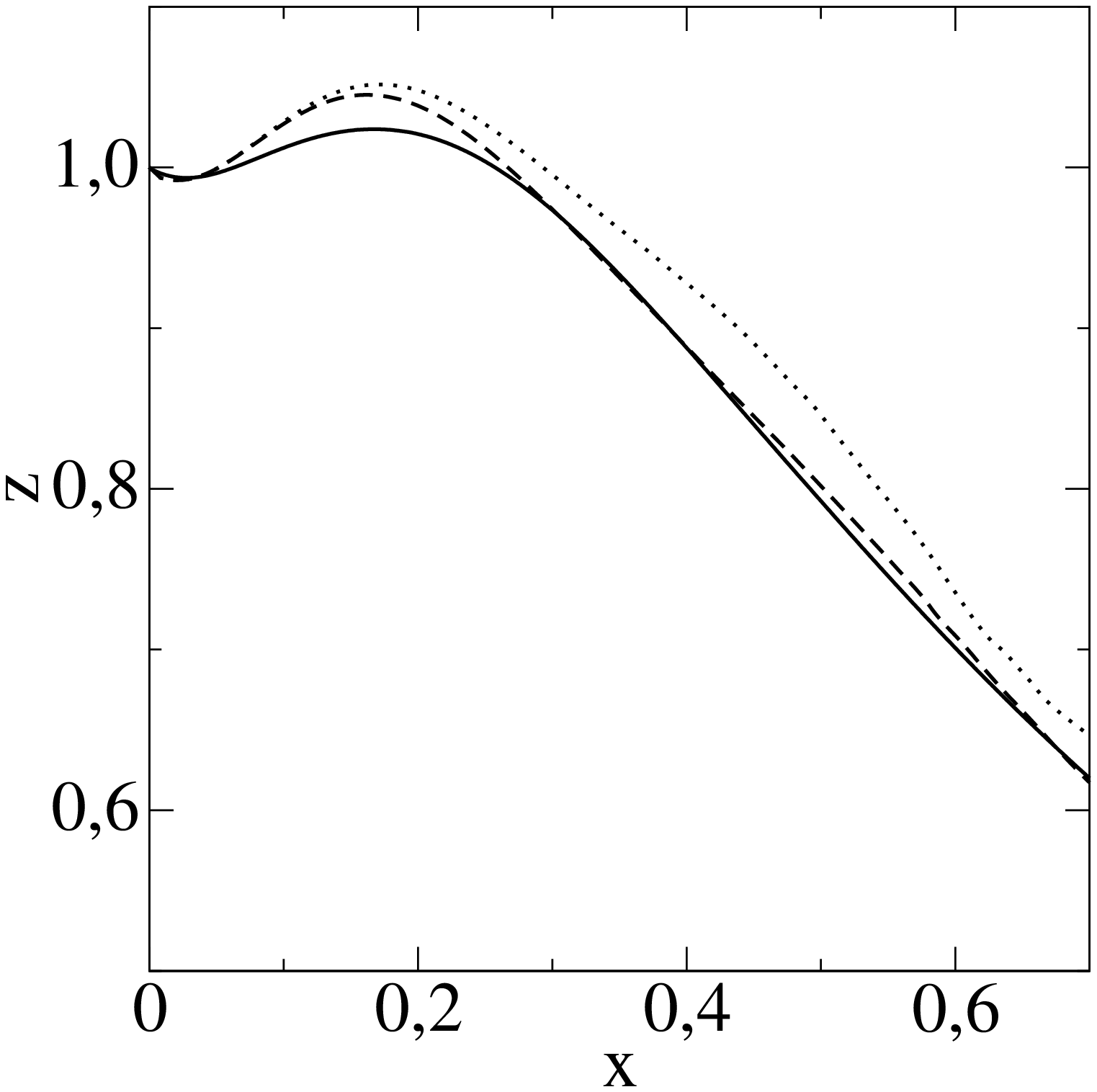}
    \caption{
Time evolution of the filtered kinetic (left) and magnetic (right) energy. DNS (solid line), LES with mixed scale similarity model using a classical dynamic procedure (dashed line) and LES with mixed scale similarity model using a divergence based dynamic procedure (dotted line) }
    \label{fig5}
\end{center}
\end{figure}
Direct and large-eddy simulations have been performed for decaying MHD turbulence. LES's use mixed-scale-similarity models both with classical and divergence-based dynamic procedures. For these LES, the computational domain is discretized using $64^3$ grid points. The filtered Navier-Stokes equation is closed by a dynamic Smagorinsky model to evaluate $\tau^{u}_{ij} - \tau^{b}_{ij}$. The influence of the SGS Lorentz force is just taken into account in the dynamic coefficient (e.g. see \cite{Muller:2002}). Figure \ref{fig5} shows the decaying of the filtered kinetic (left) and filtered magnetic (right) energy for LES and filtered DNS. The results seem to show a better agreement with the divergence-based dynamic model, but it
is difficult to be sure due to the coupling of the modeling error and an over-dissipation
of the kinetic energy in the first stage of the simulations. In further work, the modeling improvement of $\tau^{u}_{ij}$ and $\tau^{b}_{ij}$ will be addressed again using {\it a priori} tests as starting point.

\section{Conclusions}

In this work, the modeling of the subgrid-scale (SGS) term appearing in the transport
equation of the filtered magnetic field is addressed. From a priori tests, the performances
of several SGS models have been evaluated. The measure of model performance is defined as structural and functional performances. The structural performance is defined as the model capacity to locally reproduce the unknown SGS term whereas the functional performance is defined as the model capacity to reproduce the energetic action of the unknown term. The structural performance is thus evaluated by using the optimal estimation theory. This allows to compare the models but also to evaluate the improvement possibility of a given model. The functional performance is evaluated by the comparison of the GS/SGS magnetic energy transfer given by the model with the expected GS/SGS magnetic energy transfer from the DNS data. In this work, the mixed model based on the scale-similarity model with a divergence-based dynamic procedure has the best performance.
This work could be the starting point of a methodology to improve SGS modeling in various configurations. In further work, the modeling
of other SGS terms appearing in the equations of MHD flows will be addressed.
\\


The authors have benefited from fruitful discussions with CTR Summer Program participants. Computing resources were provided by IDRIS-CNRS (http://www.idris.fr/).

\vspace*{0.2in}


\end{document}